\documentclass[
 reprint,
superscriptaddress,
amsmath,amssymb,
]{revtex4-2}
\usepackage{float} 
\usepackage{graphicx}
\usepackage{dcolumn}
\usepackage{bm}

\usepackage{array}

\usepackage{hyperref}  

\makeatletter
\def\blfootnote{\gdef\@thefnmark{}\@footnotetext}
\makeatother

\begin{document}

\preprint{}

\title{Cognitive forces shape the dynamics of word usage across multiple languages} 

\author{Alejandro Pardo Pintos}
\affiliation{Department of Physics, Pabellón 1 Ciudad Universitaria, University of Buenos Aires (UBA), 1428EGA Buenos Aires, Argentina.}
\affiliation{Physics Institute of Buenos Aires (IFIBA) CONICET, Pabellón 1 Ciudad Universitaria, 1428EGA Buenos Aires, Argentina.}
\author{Diego E Shalom}
\affiliation{Department of Physics, Pabellón 1 Ciudad Universitaria, University of Buenos Aires (UBA), 1428EGA Buenos Aires, Argentina.}
\affiliation{Physics Institute of Buenos Aires (IFIBA) CONICET, Pabellón 1 Ciudad Universitaria, 1428EGA Buenos Aires, Argentina.}
\affiliation{Laboratorio de Neurociencia, Universidad Torcuato Di Tella, Buenos Aires, Argentina.}
\author{Enzo Tagliazucchi}
\affiliation{Department of Physics, Pabellón 1 Ciudad Universitaria, University of Buenos Aires (UBA), 1428EGA Buenos Aires, Argentina.}
\affiliation{Physics Institute of Buenos Aires (IFIBA) CONICET, Pabellón 1 Ciudad Universitaria, 1428EGA Buenos Aires, Argentina.}
\affiliation{Latin American Brain Health Institute (BrainLat), Universidad Adolfo Ibañez, Santiago, Chile.}
\author{Gabriel Mindlin}
\affiliation{Department of Physics, Pabellón 1 Ciudad Universitaria, University of Buenos Aires (UBA), 1428EGA Buenos Aires, Argentina.}
\affiliation{Physics Institute of Buenos Aires (IFIBA) CONICET, Pabellón 1 Ciudad Universitaria, 1428EGA Buenos Aires, Argentina.}
\author{Marcos Trevisan}
\email{marcos@df.uba.ar}
\affiliation{Department of Physics, Pabellón 1 Ciudad Universitaria, University of Buenos Aires (UBA), 1428EGA Buenos Aires, Argentina.}
\affiliation{Physics Institute of Buenos Aires (IFIBA) CONICET, Pabellón 1 Ciudad Universitaria, 1428EGA Buenos Aires, Argentina.}

\date{\today}

\begin{abstract}
\noindent The analysis of thousands of time series in different languages reveals that word usage presents oscillations with a prevalence of 16-year cycles, mounted on slowly varying trends. These components carry different information: while similar oscillatory patterns gather semantically related words, similar trends group together keywords representative of cultural and historical periods. We interpreted the regular oscillations as cycles of interest and saturation, whose behavior could be captured using a simple mathematical model. Driving the model with the empirical trends, we were able to explain word frequency traces across multiple languages throughout the last three centuries. Our results suggest that word frequency usage is poised at dynamical criticality, close to a Hopf bifurcation which signals the emergence of oscillatory dynamics. Crucially, our model explains the oscillatory synchronization observed within groups of words and provides an interpretation of this phenomenon in terms of the cultural context driving collective cognition. These findings contribute to unravel how our use of language is shaped by the interplay between human cognition and sociocultural forces.

\end{abstract}

\keywords{language usage $|$ collective cognition $|$ low dimensional model}
\maketitle

\blfootnote {Author contributions: 
A.P.P, D.E.S, G.M, and M.T. designed research;
A.P.P, D.E.S, and M.T. performed research; 
A.P.P, D.E.S, E.T. and M.T. analyzed data; 
and A.P.P, D.E.S, E.T, G.M. and M.T. wrote the paper. 
\\
The authors declare no conflict of interest.}


\noindent {\bf Significance}. The frequency with which words are used presents regular oscillations of 16 years. We propose that these oscillations arise from a basic cognitive mechanism common to other cultural objects with life cycles, such as fashion. The words that belong to a topic of interest increase their frequency, which is then inhibited by saturation until interest is regained. Here we set up a simple mathematical model for the interaction of this cognitive mechanism and the sociocultural context, which explains the occurrence frequencies of thousands of words in different languages during the past three centuries. We show that oscillations are tuned to a critical point and are synchronized within word communities.

\section*{Introduction}

Language is a superstructure in constant evolution at all levels of description. The cross-fertilization between linguistics and evolutionary biology has been enhanced by the access to massive digital corpora that now provide time series of word usage \cite{Michel2011, davies2010corpus}, opening a new era for quantitative studies in language dynamics. Joint efforts enabled to treat language dynamics using methods drawn from population genetics \cite{Newberry2017}, statistical physics \cite{Petersen2012}, and dynamical systems \cite{Abrams2003}. The study of these processes suggests that language can be understood as a system controlled by mechanisms similar to those underlying the evolution of biological species \cite{Nowak2000}. 

Word frequency is controlled by drift and selection \cite{Reali2010,Karjus2020}. Stochastic drift results from the randomness in the forms that speakers reproduce, in analogy with genetic variation \cite{Bentley2004}. On the contrary, selection is the active change of word frequency operated through imitation, memory and preferential attention to novelty \cite{Sindi2016, Amato2018}. Drift and selection have been shown to drive certain aspects of language change, including the competition of variants ({\em colour} versus {\em color}) or verb regularization ({\em lit} versus {\em lighted}). These variants evolved rather smoothly across the last two centuries, with dynamics relatively independent from the sociocultural context, and have been successfully fitted by models where imitation interacts with attention \cite{Amato2018,Shalom2019}.  

Besides cognitive forces, contextual factors can influence patterns of language use. In contrast to verb regularization and linguistic competition, noun usage is strongly associated with specific social and cultural contexts \cite{Sindi2016}. The dynamics of noun usage then result from the interplay between a largely unpredictable environment and our cognitive functions; however, the nature of this interaction remains to be elucidated. 

Previous work has already shown that word frequency evolves as regular oscillations mounted on slowly varying trends \cite{Montemurro2016}. We capitalized on this result by linking regular oscillations to the action of cognitive functions, and the trends to the larger sociocultural context. We made this operational by setting up a simple Lotka-Volterra model \cite{Lahcen2006} for collective attention and self-inhibiting saturation, in analogy with the dynamics of topic consumption in social media \cite{Lorenz-Spreen2019}. We show that our model reproduces the empirical time series and reveals that the system is tuned is to a Hopf bifurcation \cite{Strogatz01}. This means that word frequency is poised near the limit between damped and self-sustained oscillations, which is a signature of dynamical criticality \cite{Mora2011}. A theoretical prediction of our model is the appearance of collective rhythms \cite{Strogatz2000}, of which we found empirical evidence in the partial synchronization of related words.

\section*{Results} 

\begin{figure*}[ht]
  \centering
  \includegraphics[width=12cm]{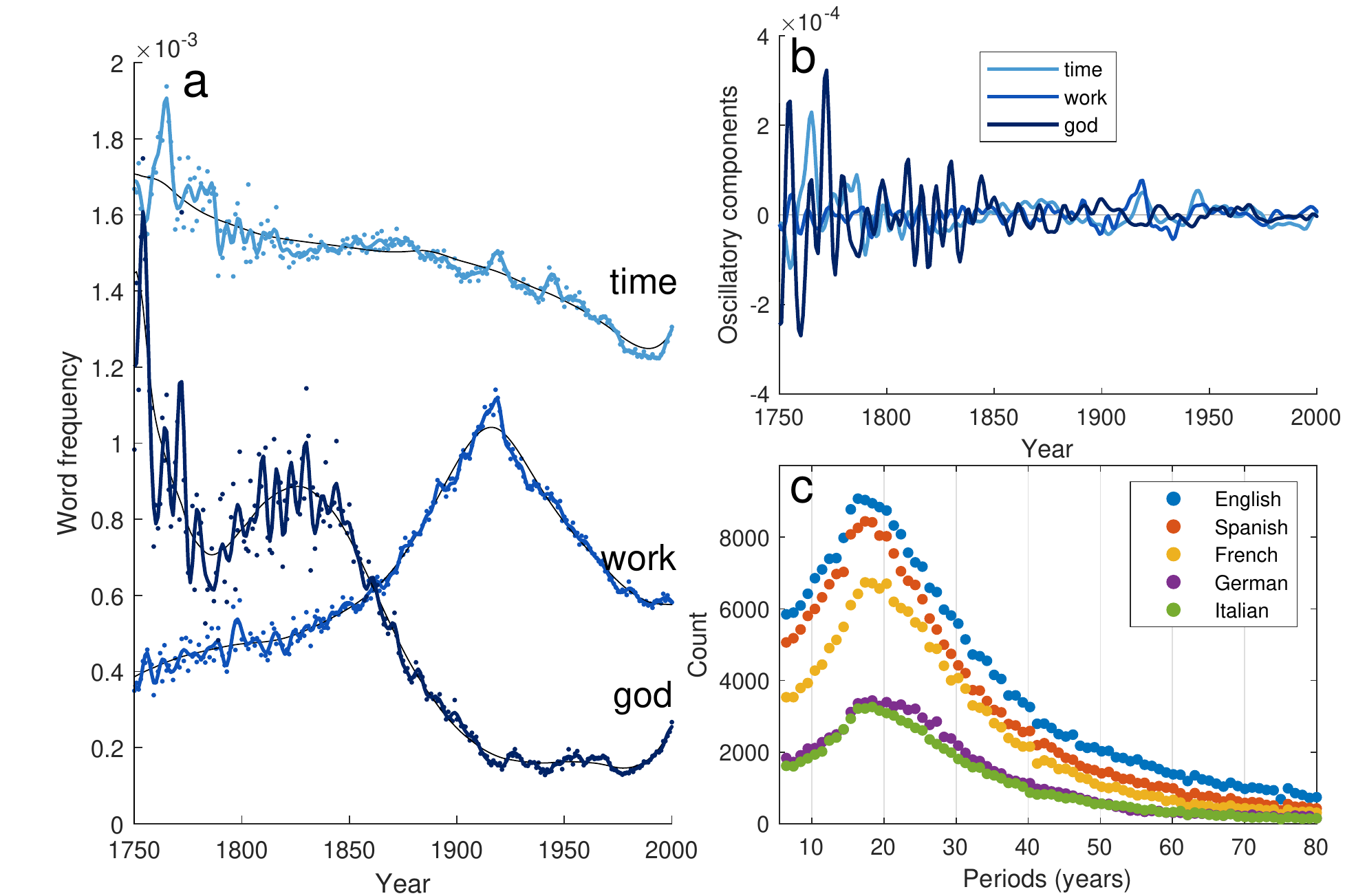}
  \caption{\textbf{Frequency of word usage presents oscillatory patterns of $\approx$16 years mounted on slowly varying trends.} (a) Points represent the frequencies $x_j(t)=N_j(t)/N(t)$ of the words {\em time, work} and {\em god}, extracted from the 2019 version of Google N-grams. Trends $tr_j(t)$ (black) were computed using singular spectrum analysis. (b) The oscillatory components were computed by subtraction $o_j(t)=x_j(t)-tr_j(t)$. (c) Following \cite{Montemurro2016}, we performed a wavelet analysis that revealed the dominance of oscillations of circa 16 years across languages.}
\label{fig:time_series}
\end{figure*}

We collected tokens of the most common nouns in English (10,403), Spanish (8,064), French (6,291), German (3,341), and Italian (2,995) from the Google N-grams 2019 data \cite{Michel2011}. For each noun $j$ we computed the frequency $x_j(t)=N_j(t)/N(t)$, with $N_j(t)$ the counts of the word $j$ and $N(t)$ the size of the corpus at year $t$. Frequency time series of the words {\em time, work} and {\em god} are shown in Figure \ref{fig:time_series}a as dots, together with the {\em trends}, i.e. non-cyclic components of the time series computed using singular spectrum analysis (shown in black). Figure \ref{fig:time_series}b shows the {\em oscillatory} components $o(t)=x(t)-tr(t)$ (blue lines). These spectrally rich oscillations show a dominance of periods around 16 years across languages, as revealed by wavelet analysis (Figure \ref{fig:time_series}c, see Supplementary Materials).

\begin{figure*}[ht]
\centering
\includegraphics[width=16cm]{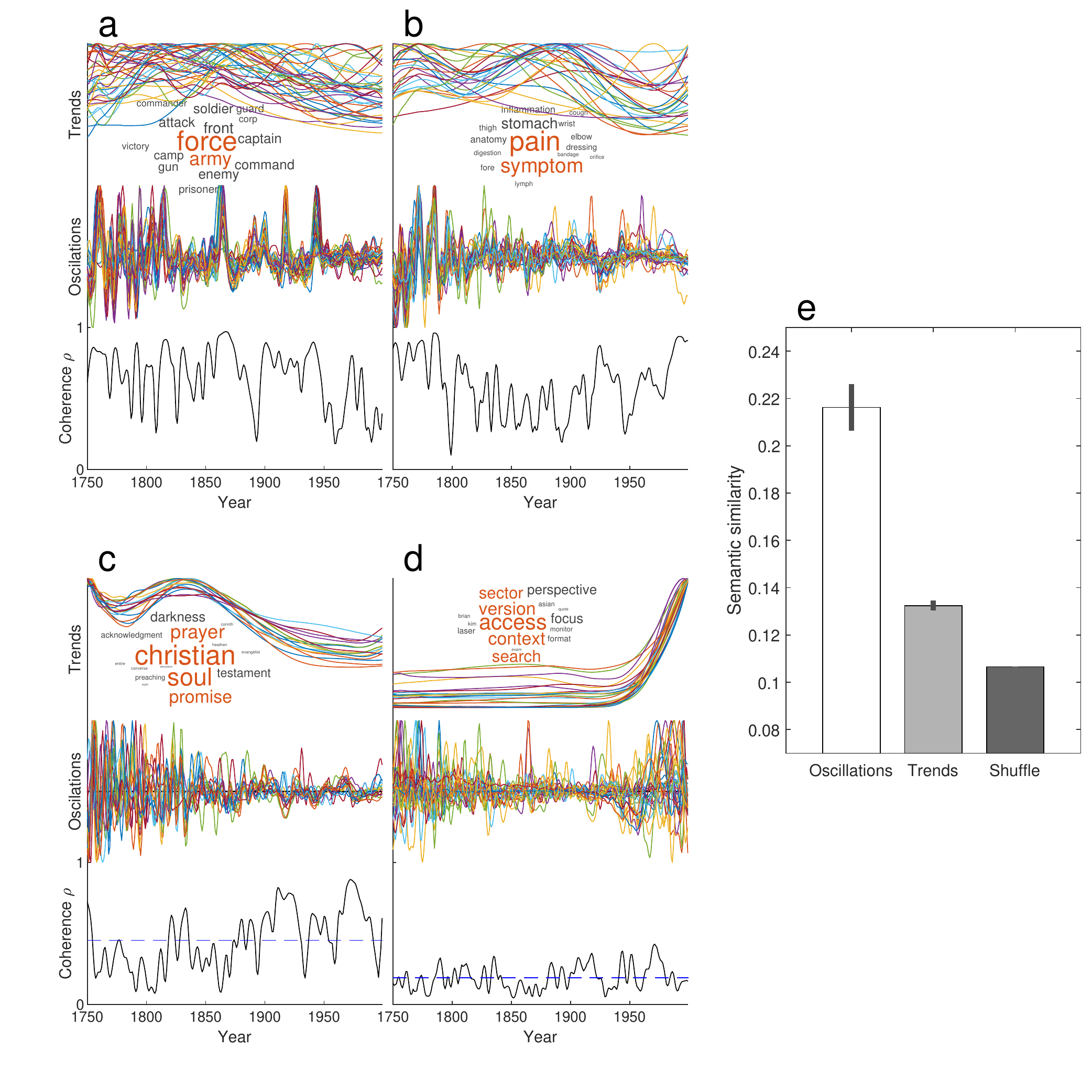}    
\caption{\textbf{Oscillations and trends carry different information.} (a-b) When only oscillations are used for clustering, communities gather words with similar semantic content. Panels correspond to the trends, oscillations and phase coherence of two oscillation clusters (time series are normalized to simplify the visualization of word frequencies across many orders of magnitude). (c-d) When only trends are used, clusters are formed that define keywords of sociocultural periods. We show trends, oscillations and phase coherence of two trend clusters. Mean coherence values are indicated with blue horizontal lines.
(e) Semantic similarity between pairs of words were computed using the fastText algorithm trained with the Wikipedia. The semantic similarity of a community was then computed as the average between the all the pairs it contains. The distributions of semantic similarities are significantly different for oscillation and trend communities, and both are higher than chance level, estimated by shuffling the words across communities.}
\label{fig:communities}
\end{figure*}

To investigate the content of the trend and oscillatory components, we classified the time series of the English corpus using hierarchical clustering with linear correlations as a similarity measure. When oscillatory components alone were used as input data, communities of semantically related words were formed. We found, for instance, a cluster formed by military-related words such as {\em force, army, gun, commander, captain, soldier, prisoner} and {\em enemy} (Fig. \ref{fig:communities}a); or a cluster related to medical terms that included the words {\em pain, symptom, inflammation, anatomy, cough} and {\em bandage} (Fig. \ref{fig:communities}b), among many others. Since we were interested in capturing persistent regularities in the oscillations along 250 years, new words coined in this period were excluded from the analysis, as well as those that fell in disuse. Furthermore, not all nouns remained semantically stable in this period \cite{Hamilton2016}. For instance, words such as {\em guy} and {\em call}, which acquired new meanings in the 19\textsuperscript{th} century, were not associated with any community in this representation. Conversely, not all the communities represented connected words; the word {\em gay}, that shifted radically in meaning in the 20\textsuperscript{th} century, belonged to a large community of rather disconnected words. Although not exhaustive, this description captures the fact that related stable nouns tend to oscillate together.

A different result is obtained when trend data is used for clustering. In this case, the groups of words that were formed tended to drift together tightly across time. For example, a cluster with high trend values during the early 19\textsuperscript{th} century that decreased afterwards included the words {\em christian, prayer, darkness, testament} and {\em promise} (Fig. \ref{fig:communities}c). Another cluster of words with a steep increase in the trends in the last decades collected the words {\em search, context, access, version} and {\em monitor} (Fig. \ref{fig:communities}d), which can all be considered keywords of the present time. {\em Monitor} is another word that changed meaning around 1930 \cite{simpson1997oxford}, in coincidence with a substantial increase in its frequency of use. Far from being an obstacle for the detection of trend communities, variations in the meaning of words can be considered signatures of the sociocultural changes \cite{Karjus2020} that characterize these clusters.  

These examples suggest that trend clusters reveal keywords specific to different sociocultural periods rather than semantically related words. To quantify this, we used a fastText word embedding model trained with the Wikipedia corpus. This neural network was trained to infer words based on their context, thus learning a low dimensional representation of the text from which a measure of semantic similarity between pairs of words could be computed. In Fig. \ref{fig:communities}e we show that the distribution of semantic similarity across communities was significantly different for trend and oscillation communities (t(287)=-10.01, p$\ll10^{-3}$). Oscillation clusters grouped words of higher semantic similarity (mean=0.22, SD=0.10) compared to trend clusters (mean=0.13, sd=0.03), and both were higher than the chance level. 

\begin{figure*}[ht]
\centering
\includegraphics[width=17cm]{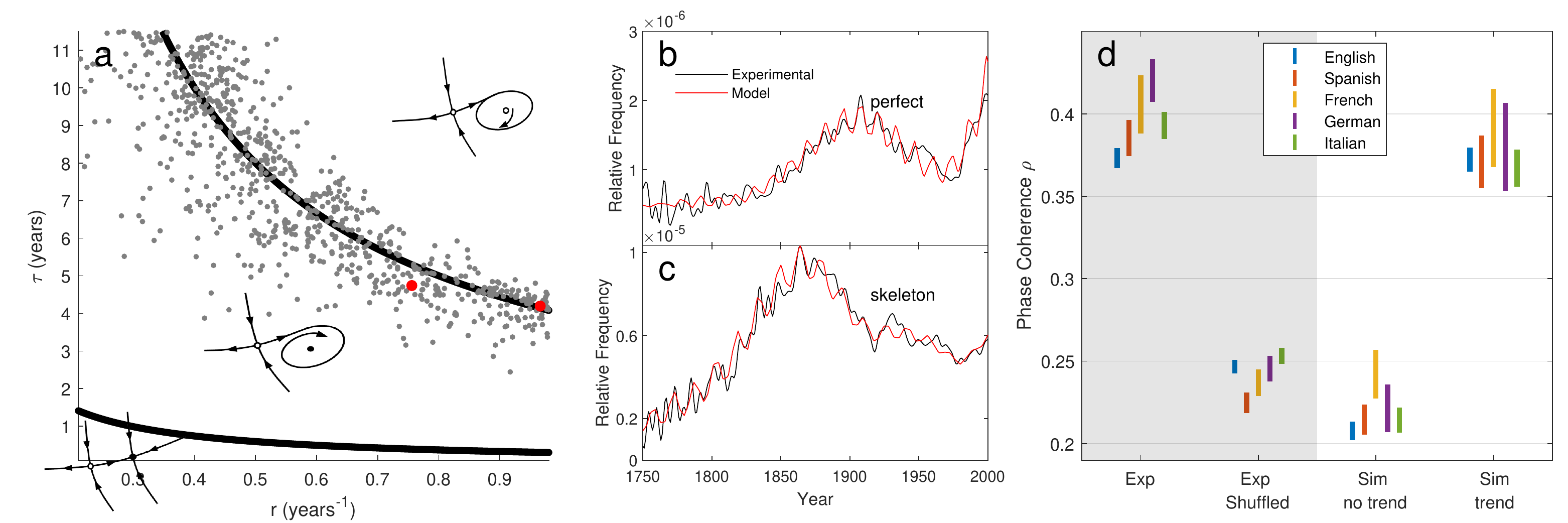}  
  \caption{\textbf{Oscillations are tuned to the Hopf bifurcation and synchronized by contextual forces.} (a) Bifurcation diagram of the 3-dimensional system of equations \ref{eq:1}. The system presents two equilibria, $x^*_{1,2}$ The origin $x^*_1$ is a saddle node, and $x^*_2$ undergoes a Hopf bifurcation at $\bar\tau=4/r$ (upper black line). Oscillations become increasingly damped until they disappear at $\bar\tau=8/27r$ (lower black line). The dimension not shown is attractive across the parameter space. English nouns fitted by the model are shown as gray points (mean growth rate $r=0.5\pm0.2$ (SD) years$^{-1}$ and mean saturation delay $\bar\tau=8\pm3$ (SD) years). (b-c) Examples of individual time series fitted with the model (red dots in panel a). (d) For different languages, we computed the distribution of mean coherence $\langle\rho\rangle$ across communities for experimental data (Exp), shuffling words across communities (Exp Shuffled), simulating words with the model at constant trend values (Sim no trend) and simulating words driving the model with the experimental trends (Sim trend). Distributions (Exp) and (Sim Trend) are equivalent.}
\label{fig:par_space}
\end{figure*}

To interpret these results, we propose a deliberately simple model with global parameters that represent cognitive factors and contextual driving. The model describes two main forces acting on the frequency $x$ (see equations in Methods). A word that belongs to a topic of interest increases its frequency of usage at rate $r$. This growth is limited by a saturation produced by the sustained consumption of the topic in the past, with a mean delay of $\bar\tau$ years. When the expansion of a word is balanced out by saturation, the frequency of use reaches an equilibrium $x^*$. This equilibrium is stable for short delays $\bar\tau<4/r$ and unstable elsewhere, as sketched in Fig. \ref{fig:par_space}a. As the average delay is increased over the critical value $\bar\tau=4/r$ (upper black curve), a stable self-sustained oscillation is created through a Hopf bifurcation. On the contrary, when the average delay is decreased and thus the relative importance of the recent past is increased, oscillations become more and more damped until they disappear at $\bar\tau=8/27r$ (lower black curve). In summary, this low dimensional model predicts the appearance of sustained oscillations as the saturation delay is increased over a threshold. 

Word usage undergoes the action of contextual forces that are intractable with low dimensional dynamical systems. Here we assume that the trend $tr(t)$ represents the resultant of these forces on a word, which drives the equilibrium frequency $ x^*=tr(t)$, leaving the growth rate $r$ and the delay $\bar\tau$ as free cognitive parameters to fit the experimental time traces. In Fig. \ref{fig:par_space}b and c we show examples of word frequency data (black) successfully reproduced by the model (red). Interestingly, the fitting parameters were distributed along the Hopf bifurcation, shown as gray dots in Fig. \ref{fig:par_space}a, presenting a mean imitation rate of $r=0.5\pm0.2$ years$^{-1}$ and mean saturation delay $\bar\tau=8\pm3$ years. 

 Fitting the model to empirical data allowed us to build a picture of word frequencies behaving as near to self-sustained oscillators while driven by sociocultural forces. Beyond providing a plausible interpretation for word frequency dynamics, the model yielded a specific prediction: if word usage arises from a mechanism of driven oscillators, then words with similar drives would partially synchronize their cycles. Since trend communities are formed precisely by words that share similar drives, we expect to observe some degree of synchronization in the words of these communities. To quantify the collective rhythm of a community, we computed the phase coherence $\rho$. When $\rho\sim 0$ the oscillators have random phases, while $\rho=1$ means that the oscillators are collectively in phase. This happens, for example, when the maxima or the minima of all the oscillations coincide at the same time, and thus the cluster is completely synchronized.

Typical traces of $\rho(t)$ for trend clusters are shown in the lower panels of Figs. \ref{fig:communities}c and \ref{fig:communities}d, along with the mean coherence values $\langle\rho\rangle$ marked by blue horizontal lines. These non-zero coherence values may indicate an effect of finite size; in fact, small communities of unforced oscillators show some degree of coherence \cite{Strogatz2000}. To investigate this possibility, we first computed the distribution of mean coherence values across communities for different languages (Fig. \ref{fig:par_space}d Exp). Then we shuffled the words between communities and recomputed the coherence, obtaining much lower values (Fig. \ref{fig:par_space}d Exp Shuffled). This supports that oscillations are indeed partially synchronized within the communities (for further testing, see Methods).

To explore the sources of this coherence within the framework of our model, we simulated every community integrating our model with parameters $(r,\bar\tau)$ distributed near the Hopf bifurcation and with random initial conditions. When simulations were performed using constant trend values, the coherence levels were similar to those of the shuffled population (Fig. \ref{fig:par_space}d Sim no trend). Instead, when equations were driven by the empirical trends, the coherence increased (Fig. \ref{fig:par_space}d Sim trend) to the levels observed experimentally in all the tested languages. This shows that the trends act as a driving that contributes to synchronize the word frequencies, providing the empirically observed coherence values between words within communities.

\section*{Discussion}

The analysis of thousands of time series corresponding to different languages revealed that word frequencies present spectrally rich content, with a dominance of 16-year oscillations. We propose that these regular oscillations arise from a mechanism common to the life cycles of many cultural objects \cite{Bikhchandani1992,Sproles1981}, which is particularly apparent in the periodic comebacks of fashion styles. In language, cycles are characterized by a growth in the frequency of words that form a topic of interest, which is then inhibited by saturation until interest is regained. We made this mechanism operational with a simple model that increases word frequency of use due to attention and decreases it by saturation. Of course, attention is associated with topics rather than individual words. However, topical fluctuations explain a significant amount of variability in the change of individual word frequencies \cite{Karjus2020, Sindi2016}, which we used as variables in our model.

The model is completed by driving this cognitive-based mechanism with the empirical trends, which we associated with the specifics of sociocultural contexts. This context represents economic events, natural disasters and wars \cite{Montemurro2016} but also subjective factors as life-styles and well-being \cite{Hills2019}, as well as other attractors of collective interest that alter word frequency. 

Some caution is needed regarding the treatment of cognitive and sociocultural forces as distinct from one another. This separation is not straightforward, as sociocultural forces affect cognitive processes, particularly the ones related to attention. This is indeed the case for the consumption of popular content that exhausts the attention resources more and more rapidly over time \cite{Lorenz-Spreen2019}. In our model, this relation between cognition and culture could be accounted by the use of time-dependent parameters. However, here we kept the simplest model capable of explaining the strong regularities in word usage across centuries in different languages.

As simple as it is, the model allowed us to reproduce global features of word use traces. Parameter fitting indicated that the maximum influence on the growth rate response is due to the consumption of the words 5-11 years in the past. This can be seen as a  timescale related to `social interest', much in the same way that Michel and collaborators \cite{Michel2011} analyzed the decreasing levels of interest in events of the past and showed that, in recent years, interest decline to half of the initial level after only 10 years. 

Beyond the cognitive factors that regulate word cycles, a number of cultural and biological forces are involved in language change at different levels, such as individual learning \cite{Reali2010, Kirby2015,Tamariz2016}, changes in vocal anatomy \cite{Blasi2019} and adaptations that decrease the effort of language production and understanding \cite{Mahowald2018, Kanwal2017}, among others. Decreased effort refers to the optimization of the trade-off between accurate and efficient communication, which has been put forward as a possible explanation of Zipf's law \cite{FerreriCancho2003}. This empirical law states that the frequencies of the words used in any text are ordered in terms of an universal power law as a function of their rank \cite{Piantadosi2014}, a signature of statistical criticality. Interestingly, we found that word frequencies are auto-organized in a narrow band around the Hopf bifurcation, close to the limit between damped and self-sustained oscillations. This tuning to a Hopf bifurcation is a signature of dynamical criticality that has been observed in biological systems \cite{Eguiluz2000}. Except for a few cases, the mathematical treatment used to describe criticality in statistical systems is quite different from the one used for dynamical systems \cite{Mora2011}. Since language is dynamical by nature, we believe that this finding could provide a clue to address possible relationship between statistical and dynamical notions of criticality in word frequency time series. 

We have described the global dynamics of word usage in different languages using a basic attention-saturation mechanism that produces cycles. Beyond providing a plausible interpretation of word frequency data, we believe that the model raises two interesting theoretical implications. First, the notion of dynamical criticality in word frequency time series, and its possible relationship with universal aspects of language. Second, the dynamical structure of forced oscillations predicts the appearance of partial synchronization, which is indeed observed in communities formed with words of similar trends. We expect that these findings will open an avenue in the investigation of language dynamics and contribute to unravel how human cognition and sociocultural forces interact to shape our use of language.

\section*{Materials and Methods}
\label{sec:Methods}

\subsection*{Corpus and data processing}
Occurrence frequencies $x_j(t)=N_j(t)/N(t)$ were computed using the counts $N_j(t)$ of the word $j$ and the number of words in the corpus $N(t)$ at each year $t$. We used singular spectral analysis (SSA) \cite{Zhigljavsky2011} to extract cyclic and non-cyclic components from the time series $x(t)$. Trends $tr(t)$ were computed using the non-cyclic components of each word using the Matlab function \texttt{autotrend} \cite{alexandrov2005automatic}. 

The Google Books is a massive corpus of lexical data extracted from $\sim$ 8 million books (6\% of all books ever published) that has been widely used for research. Despite of its size, is not free from biases \cite{Pechenick2015}, which we addressed as follows: 

\begin{itemize}
\item {\em Uneven topic representation}. An unbalance in the sampling of topics has been reported for the English corpus of the Google Books 2012 \cite{Pechenick2015}. Following the suggestions in \cite{Younes2019}, we performed our analyses on different language corpora. Nouns were extracted from the latest Google 1-gram 2019 database, and converted to singular form \cite{Bird2009}. To increase the statistical power, we kept nouns present every year and at least $10^6$ times within the interval 1750–2000. This left us with a core vocabulary of 10,403 English, 8,064 Spanish, 6,291 French, 3,341 German, and 2,995 Italian nouns. 

\item {\em Random sampling}. To avoid random sampling effects in the database loading \cite{Karjus2020}, the oscillatory components $o_j(t)=x_j(t)-tr_j(t)$ were low-pass filtered, keeping the frequencies $f<1/6$ years$^{-1}$. The rationale behind this is supplied in the Supplementary Materials, and we illustrate it here with an example. Consider for instance the amount of religious books loaded to the database in successive years. If this number varies from one year to the other, religious words that appear together in those books will present time traces highly correlated in the frequency of the database loading, $f\sim 1$ year$^{-1}$. To avoid this bias, high frequencies were removed by low-pass filtering the time series (Figure \ref{fig:time_series}a and b).

\end{itemize}

\subsection*{Clustering}
We computed the hierarchical cluster tree using linear correlations as a similarity measure using the MATLAB function \texttt{linkage} on the frequencies $x(t)$, trends $tr(t)$ and oscillations $o(t)=x(t)-tr(t)$ across the period (1750-2000). Clusters from $x(t)$ and $tr(t)$ produce virtually the same word groups. The cutoff levels were set to 0.04 for trends and 0.5 for oscillations; these  values are the lowest possible that ensure maximum correlation between series compatible with a cluster size distribution that follows the Zipf's law. Communities of less than 10 words were discarded. Codes are supplied to reproduce the complete set of trend and oscillation clusters.

\subsection*{Semantic similarity}
The fastText model trained with the English Wikipedia database was used to estimate the mean semantic similarity within each community \cite{Bojanowski2017}. Semantic similarity is higher for oscillation clusters  than for trend clusters (oscillations: mean=0.22, SD=0.10; trends: mean=0.13, SD=0.03; t(287)=-10.015, p<$10^{-19}$). Chance levels were computed by shuffling the words across all communities. Semantic similarity of both types of clusters is significantly higher than chance (oscillations: t(226)=11.1026, p<$10^{-22}$; trends: t(384)=12.4769, p<$10^{-29}$).

\subsection*{Model} 
The single species model reads $\dot x(t)=r\,x(t)$, where $x(t)$ represents the unit density of population at time $t$, and $r>0$ is the intrinsic rate of growth for population \cite{Ladas1994}. Considering the competition for finite resources, the equation becomes $\dot x(t)=r x(t)[1-x(t)/x^*]$, where $x^*>0$ is the non-zero equilibrium population. When time delay becomes important, the system is governed by $\dot x(t)=r x(t)\left[1-{{x(t-\tau)}/x^*}\right]$. Here we used 

\begin{equation}
\label{eq:1}
\dot x(t)=r x(t)\left[1-{\frac{1}{x^*}} \int_{-\infty}^t G(t-\tau)x(\tau)d\tau\right],
\end{equation}

\noindent where $G(t)$, called the delay kernel, is a weighting factor that indicates how much emphasis should be given to the frequency $x$ at earlier times to determine its effect in the present. Here we used the strong kernel $G(\tau)=4\tau/\bar\tau^2\ e^{-2\tau/\bar\tau}$, which increases from zero to a maximum at $\bar\tau/2$ and then decays exponentially. This functional form assumes that there is a preferential delay for the influence of the past, with an average of $\bar\tau=\int_0^\infty uG(u)\ du$. 

Defining the integrals $y(t)=2/\bar\tau\int_{-\infty}^t e^{-2(t-s)/\bar\tau}x(s)\, ds$ and $z(t)=4/\bar\tau^2\int_{-\infty}^t (t-s) e^{-2(t-s)/\bar\tau}x(s)\, ds$, and applying the chain rule \cite{Lahcen2006}, equations \ref{eq:1} can be further reduced to the system 

\begin{equation} \label{eq:2}
    \left\{
        \begin{array}{ll}
          \dot x=rx\left[1-z/x^*\right]\\
          \dot y=2/\bar\tau (x-y)\\
          \dot z=2/\bar\tau (y-z).
        \end{array}
    \right.
\end{equation}

Equations \ref{eq:2} have two equilibria, a saddle node at the origin $x^*_1=0$, and another at $x^*_2=x^*$. Linearization at this non-zero equilibrium gives the characteristic equation $\lambda^3+4/\bar\tau\,\lambda^2+4/\bar\tau^2\, \lambda+4r/\bar\tau^2=0$. When $\bar\tau<4/r$, all three roots of the characteristic equation have negative real part and the equilibrium $x^*$ is stable; at $\bar\tau=4/r$, the characteristic equation has a negative real root $\lambda_1=-r$ and a pair of purely imaginary roots $\lambda_{2,3}=\pm ir/2=$, giving rise to self-sustained oscillations of frequency $\omega=r/2$ through a Hopf bifurcation shown in Fig. \ref{fig:par_space}a (upper black curve).
As the average delay is increased over the critical value $\bar\tau=4/r$, a stable periodic solution is created. When on the contrary the average delay is decreased, making the recent past more important, oscillations become more and more damped until they disappear at $\bar\tau=\frac{8}{27r}$, where the three roots of the characteristic equation are real (lower black curve in Fig. \ref{fig:par_space}a). In summary, the Lotka-Volterra system with strong kernel is a low dimensional model that predicts the appearance of sustained oscillations as the saturation delay is increased over a threshold.

\subsection*{Numerical integration and parameter fitting}

Our model was used to simulate individual words $x_m(t)$ by integration of equations \ref{eq:2} driven by the experimental trends $x^*=tr(t)$, and initial conditions $(x_0,y_0,z_0)=(tr(1750),tr(1750),tr(1750))$. 
We ran the model for every point in the grid $0.2\leq r\leq 1$ ($\delta r=0.01$) and $1\leq \bar\tau\leq 12$ ($\delta \bar\tau=0.1$) of Fig. \ref{fig:par_space}a and selected the simulation minimizing the objective function $zscore(log(1-corr_{osc}))+zscore(log(D))$, a mixture of the total difference $D=\left[\int[x_m(t)-x_e(t)]^2 dt\right]^{1/2}$ and the correlation between the oscillations $o_e(t)$ and $o_m(t)$. 

Fitting of individual words involves comparing experimental traces $x_e(t)$ of slightly variable frequency with simulations $x_m(t)$ of fixed frequency. Due to the difficulty of constructing accurate similarity measures between such traces, we selected the 996 
words for which the oscillations of simulations and experimental traces have similar amplitude: $3/4 \, SD(o_e(t)) < SD(o_m(t)) < 5/4 \, SD(o_e(t))$.    

\subsection*{Phase coherence}
We transformed the oscillatory components $o_j(t)$ to phase variables $\theta_j(t)$ using the Hilbert transform \cite{Matsuda2017}. The collective rhythm of a community was computed with the order parameter $\rho e^{i\psi}=\sum_{j=1}^N e^{\theta_j}/N$, summing over the $N$ words that form the community. This complex number quantifies the collective rhythm produced by the population of words within a community, where $\rho(t)$ measures the phase coherence and $\psi(t)$ the phase average of a community.

Figure \ref{fig:par_space}d shows the distribution of the mean coherence values $\langle\rho\rangle$ across communities for the following conditions: 1. the experimental communities (Exp), 2. shuffling words between communities (Exp Shuffle), 3. Simulations of word communities using the model with constant trend values (Sim no trend), 4. Simulations of word communities driving the model with the experimental trends (Sim trend). Coherence values are normally distributed (Kolmogorov-Smirnov test: $p>0.05$ for all languages and conditions). A two-sample t-test (Table \ref{tab:ttest}) shows that only the experimental (Exp) and simulated with trends (Sim trend) distributions are equivalent across languages.

\subsection*{Data and code availability}
All the datasets are publicly available at \href{http://storage.googleapis.com/books/ngrams/books/datasetsv2.html}{http://storage.googleapis.com/books/ngrams/books/ datasetsv2.html}. The codes and processed data used to generate the figures of this work and the word communities are available in matlab at \href{https://github.com/AlePardoPintos/Cognitive-forces-words}{https://github.com/AlePardoPintos/Cognitive-forces-words}


\begin{table}[H]
\centering
\resizebox{0.8\columnwidth}{!}{
\begin{tabular}[c]{|c|c|c|c|}
\hline
 & Exp vs & Exp vs & Exp vs\\
 & Exp Shuffled & Sim No Trend & Sim Trend \\
\hline
English & t(378)=17.001, & t(378)=19.730, &  t(378)=0.086,\\
& p$<10^{-20}$& p$<10^{-20}$& p$>$0.05\\
\hline
Spanish & t(219)=14.562, & t(219)=15.883, & t(219)=1.853, \\
& p$<10^{-20}$& p$<10^{-20}$& p$>$0.05\\
\hline
French & t(146)=12.717, & t(146)=11.975, & t(146)=0.737, \\
& p$<10^{-20}$& p$<10^{-20}$& p$>$0.05\\
\hline
German & t(60)=8.702, & t(60)=7.107, & t(60)=0.483, \\
& p$<10^{-12}$& p$<10^{-9}$& p$>$0.05\\
\hline
Italian & t(58)=11.567, & t(58)=10.272, & t(58)=1.350, \\
& p$<10^{-16}$& p$<10^{-14}$& p$>$0.05\\
\hline
\end{tabular}
}
\caption{\footnotesize{Two-sample t-test for comparing coherence levels between experimental data and simulations (see Fig. \ref{fig:par_space}d).}}
\label{tab:ttest}
\end{table}

\begin{acknowledgments}
The research in this work was partially funded by the Consejo Nacional de Investigaciones Científicas y Técnicas (CONICET), the University of Buenos Aires (UBA).
\end{acknowledgments}

\bibliography{MyCollection}

\end{document}



\maketitle


\section*{Supplementary Information}

\subsection*{Spectral analyses}
\begin{figure}
\centering
\includegraphics[width=11cm]{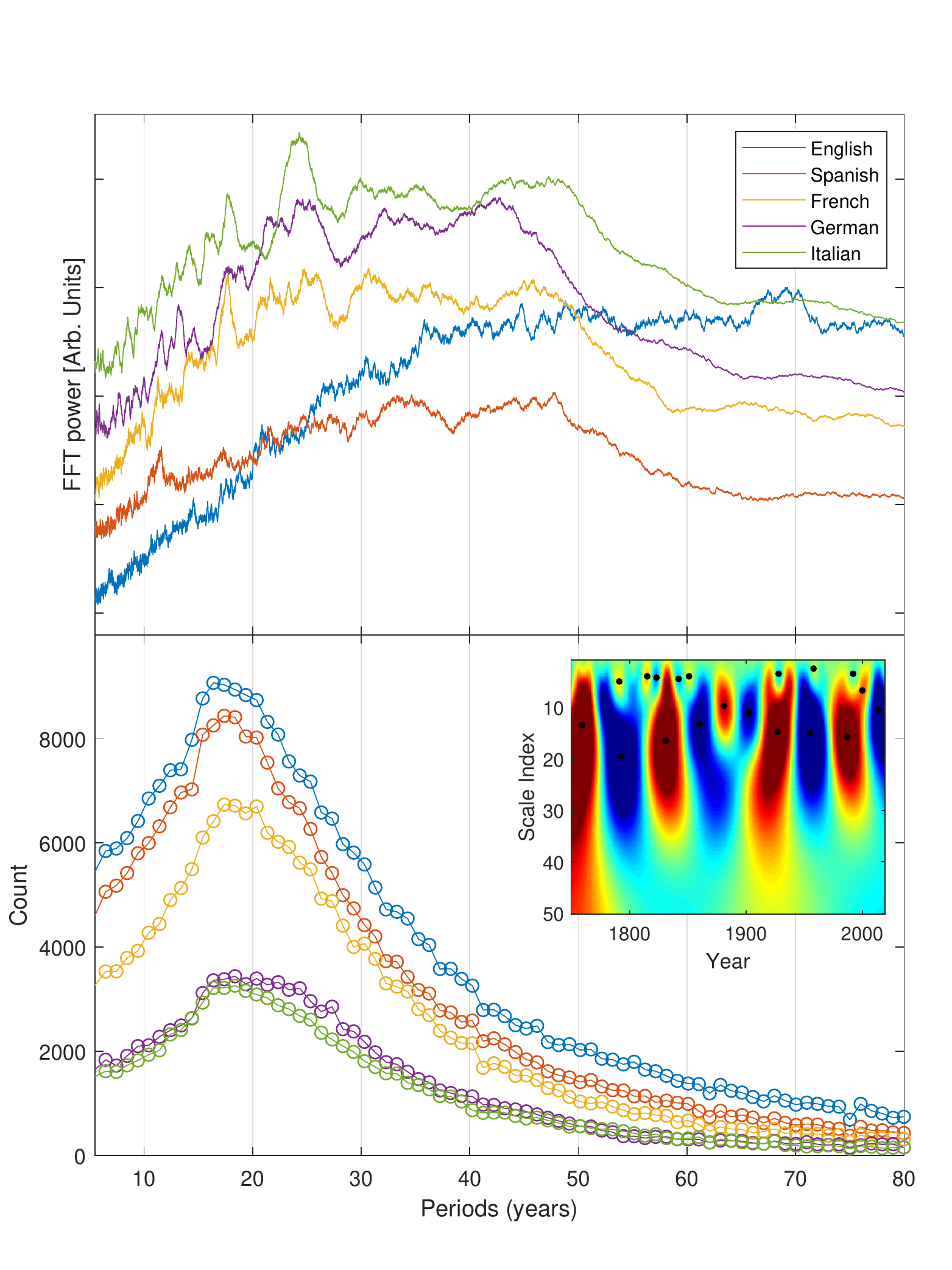}
\caption{Spectral analyses. Upper panel: Fast Fourier analysis shows local peaks for oscillatory components with periods between 10 and 30 years across languages. Lower panel: scalograms were computed for individual words (inset), averaging their maxima across corpora. The histogram reveals a dominance of 16-years cycles.}
\label{fig:s1}
\end{figure}

Time series of word usage $x(t)$ were decomposed into trend $tr(t)$ and oscillatory $o(t)=x(t)-tr(t)$ components. To characterize the oscillations, we computed the Fast Fourier Transform  $o_i(t)$ and then averaged across words $i$. The result is shown in the upper panel of Figure \ref{fig:s1} for English, Spanish, French, German and Italian. 
This analysis revealed that the oscillations are spectrally rich, with local peaks for periods between 10-30 years for the different languages. 

Given the non-stationary nature of our signals, we performed a wavelet analysis to capture the intervals of characteristic oscillations. Following the procedure described in \cite{Montemurro2017}, we computed extrema of wavelet coefficients from the scalograms of each word (Figure \ref{fig:s1}, inset) averaged across words. The result is shown in the bottom panel of Figure \ref{fig:s1}, revealing a dominance of $\sim$ 16 years oscillations for all languages. 

\subsection*{Separating random and systematic effects}

\begin{figure}
\centering
\includegraphics[width=15cm]{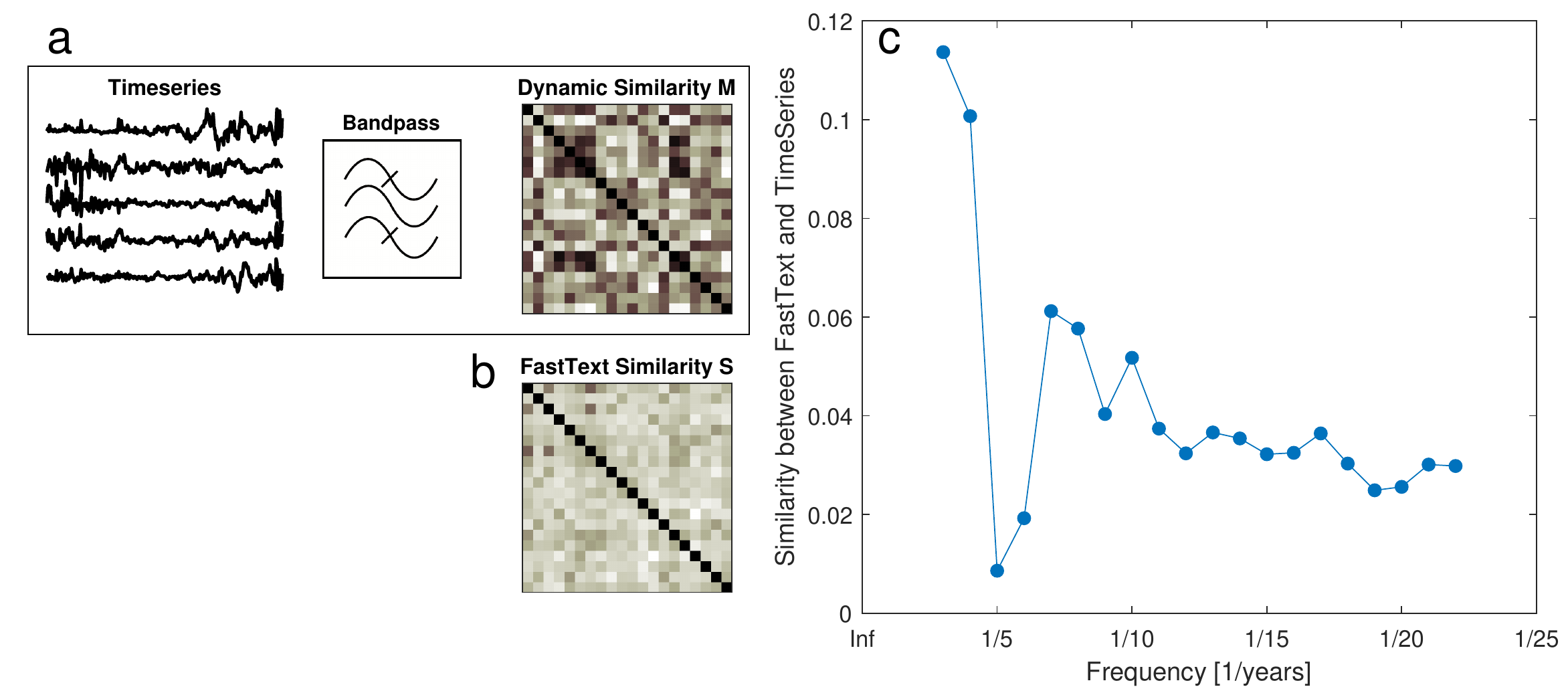}
\caption{Oscillatory components transport semantic content at different frequencies. (a) Band-pass filters were applied to select oscillatory components of each word at different frequencies; the similarity matrix  $M^f$ measures the correlation between the oscillatory components for each pair of words in the corpus. (b) The sematic similarity $S$ was computed training the FastText word embedding model with the Wikipedia. (c) Correlation between matrices $M^f$ and $S$ was computed. Semantic information is high for rapid oscillations ($f>1/4$ years$^{-1}$) and for slow oscillations ($f<1/7$ years$^{-1}$). The same behavior was obtained for different bandwith values, $1/5<\Delta f<1$.}
\label{fig:s2}
\end{figure}

We investigated the information conveyed by the oscillations $o_i(t)$ at different frequencies. For that sake, we:

\begin{itemize}
    \item Applied a band-pass filter $p_f(o_i)$ (Matlab function \texttt {bandpass}), which selects the oscillatory components around the frequency $f$ of word $i$. We then computed the similarity matrix $M^f_{ij}=cor(b_f(o_i),b_f(o_j))$. This matrix tells us how similar are the oscillations of words $i$ and $j$ around the frequency $f$ (Figure \ref{fig:s2}a). 

    \item Computed the semantic similarity matrix $S_{ij}$ between words $i$ and $j$ using the fastText word embedding model \cite{Bojanowski2017} trained with the independet corpus of Wikipedia. FastText is a neural network trained to infer words based on their context, thus learning a low dimensional representation of the text from which a measure of semantic similarity $S_{ij}$ between pairs of words $i$ and $j$ can be derived (Figure \ref{fig:s2}b). 

\end{itemize} 

Comparing the matrices $M^f$ with $S$ tells us how much semantic information is conveyed by the oscillatory components of frequency $f$. The results are shown in Figure \ref{fig:s2}c, where the y-axis is the correlation between matrices $M^f$ and $S$, and the x-axis represents the central frequency $f$ of the filter $b_f$.

Figure \ref{fig:s2}c shows that the semantic information is high for rapid oscillations ($f>1/4$ years$^{-1}$) and for slow oscillations ($f<1/7$ years$^{-1}$). The high semantic content found for rapid oscillations is due to a bias in the database loading. To understand this, consider for instance the variations in the amount of religious books loaded to the database in consecutive years. As a result of this random loading, religious words that appear together in those books will present time traces highly correlated in the frequency of the database loading, $f\sim 1$ year$^{-1}$. 

Interestingly, semantic content rises for slower oscillations $f<1/7$ years$^{-1}$, which is not attributable to biases in the database. Together with the fact that words present a dominance of oscillations with frequency $f\sim 1/16$ years$^{-1}$, this helps building confidence that slow oscillations express the action of collective cognitive factors on word usage.  

\bibliographystyle{apsrev4-2}
\bibliography{MyCollection}